\documentclass[aps,pra,preprint,showpacs,showkeys,groupedaddress]{revtex4-1}
\usepackage{amsmath}
\usepackage{amsfonts}
\usepackage{amssymb}
\usepackage{amsthm}
\begin{document}


\title{Bound entanglement and distillability of multipartite quantum systems}



\author{Hui Zhao$^\dag$  Xin-yu Yu $^\dag$ Naihuan Jing $^\ddagger$ }
\affiliation{$^\dag$ College of Applied Science, Beijing University
of Technology, Beijing 100124, China}

\affiliation{$^\ddagger$ Department of Mathematics, North Carolina State University, Raleigh, NC 27695, USA\\
$^\ddagger$ School of Mathematical Sciences, South China University of Technology, Guangzhou 510640, China}

\date{\today}

\begin{abstract}
We construct a class of entangled states in $\mathcal{H}=\mathcal{H}_{A}\otimes\mathcal{H}_{B}\otimes\mathcal{H}_{C}$ quantum systems with $dim\mathcal{H}_{A}=dim\mathcal{H}_{B}=dim\mathcal{H}_{C}=2$ and classify those states with
respect to their distillability properties. The states are bound entanglement for the bipartite split$(AB)-C$.
The states are NPT entanglement and $1$-copy undistillable for the bipartite splits $A-(BC)$ and $B-(AC)$.
Moreover, we generalize the results of $2\otimes2\otimes2$ systems
to the case of $2n\otimes 2n\otimes2n$ systems.

\end{abstract}



\maketitle

\section{INTRODUCTION}
Quantum entanglement is one of the most astonishing quantum phenomena. It plays an important
role in quantum information such as dense coding \cite{1}, quantum teleportation \cite{2}
and quantum cryptographic schemes \cite{3,4,5}.

Namely we say that a state of composite systems is considered to be entangled if it can not be written
as a convex combination of product states \cite{6}. Considerable efforts have been devoted to analyze the
separability and entanglement \cite{t1,t2,t3,t4,t5,t6}. Indeed there are two kinds of entangled states. One
is the free entangled state which is distillable, and the other is the bound entangled state. A bound
entangled state is one which is entangled and does not violate Peres condition \cite{x1}. For $2\otimes4$ and
$3\otimes3$ systems, explicit examples of bound entangled states have been constructed in Ref. \cite{j1}. It has
been shown that any state with PPT---positive partial transpose is non-distillable and a bipartite state
is distillable if some number of copies $\rho^{\otimes n}$ can be projected to obtain a two-qubit state with NPT
(non-PPT) \cite{x2}. Therefore the bound entanglement can not be brought to the singlet form by means of local
quantum operations and classical communication from many copies of a given state. Instead, is an NPT state distillable?
It was proved that for two-qubit systems all entangled states are distillable \cite{x3}. That means there is no NPT
bound entangled state in $2\otimes2$ systems. For higher dimensions, the existence of bound entangled state
with NPT has been discussed in Refs. \cite{j2,j3,j4,j5,j6,j7}.
As a matter of fact, bound entanglement can not be used alone for quantum communication, nevertheless, it
can be distillable through free entanglement \cite{j8}. Moreover, in Ref. \cite{j9} for some bound entangled
state $\rho_{1}$ with NPT there exists another bound entangled state $\rho_{2}$ such that the joint state
$\rho_{1}\otimes \rho_{2}$ is no longer a bound entangled state. Such a phenomenon is called superactivation.

In this paper, we analyze a class of tripartite entangled states. The paper is organized as
follows. In Section $2$, first we construct certain entangled states, then we give a detailed description about the entanglement with respect to different bipartite splits in $2\otimes2\otimes2$ systems. In Section 3, we generalize these results to $2n\otimes 2n\otimes 2n$ systems. Finally, conclusion and discussion are given in Section $4$.

\section{Entanglement of $2\otimes2\otimes2$ quantum systems}

In this section we consider the entanglement of mixed states for different bipartite splits in $2\otimes2\otimes2$
systems. Consider the Hilbert space $\mathcal{H}=\mathcal{H}_{A}\otimes\mathcal{H}_{B}\otimes\mathcal{H}_{C}$, $dim\mathcal{H}_{A}=dim\mathcal{H}_{B}=dim\mathcal{H}_{C}=2$. Let $P_{\phi}=|\phi\rangle\langle\phi|$, ${e_{i}}$
stand for orthonormal basis of $\mathcal{C}^{2}$, $i=1,2$. We define the vectors
\begin{eqnarray}
  \Psi_{1} &=& \frac{1}{\sqrt{2}}(e_{1}\otimes e_{1}\otimes e_{1}+e_{2}\otimes e_{1}\otimes e_{2}), \nonumber \\
  \Psi_{2} &=& \frac{1}{\sqrt{2}}(e_{1}\otimes e_{1}\otimes e_{2}+e_{2}\otimes e_{2}\otimes e_{1}), \nonumber \\
  \Psi_{3} &=& \frac{1}{\sqrt{2}}(e_{1}\otimes e_{2}\otimes e_{1}+e_{2}\otimes e_{2}\otimes e_{2}),\nonumber \\
    \Phi_{b}&=&e_{2}\otimes e_{1}\otimes(\sqrt{\frac{1+b}{2}}e_{1}+\sqrt{\frac{1-b}{2}}e_{2}), \ \ b\in[0,1].
\end{eqnarray}
We construct a state as following
\begin{equation}\label{}
    \sigma_{insep}=\frac{2}{7}\sum^{3}_{i=1}P_{\Psi_{i}}+\frac{1}{7}P_{e_{1}\otimes e_{2}\otimes e_{2}},
\end{equation}
which is inseparable for all bipartite splits. It can be verified by using the partial transposition criterion.
Now we define the following state
\begin{equation}\label{}
    \sigma_{b}=\frac{7b}{7b+1}\sigma_{insep}+\frac{1}{7b+1}P_{\Phi_{b}}.
\end{equation}
Its matrix is of the form
\begin{equation}\label{}
   \sigma_{b}= \frac{1}{7b+1}\left(
  \begin{array}{cccccccc}
    b & 0 & 0 & 0 & 0 & b & 0 & 0 \\
    0 & b & 0 & 0 & 0 & 0 & b & 0 \\
    0 & 0 & b & 0 & 0 & 0 & 0 & b \\
    0 & 0 & 0 & b & 0 & 0 & 0 & 0 \\
    0 & 0 & 0 & 0 & \frac{1+b}{2} & \frac{\sqrt{1-b^{2}}}{2} & 0 & 0 \\
    b & 0 & 0 & 0 & \frac{\sqrt{1-b^{2}}}{2} & \frac{1-b}{2} & 0 & 0 \\
    0 & b & 0 & 0 & 0 & 0 & b & 0 \\
    0 & 0 & b & 0 & 0 & 0 & 0 & b \\
  \end{array}
\right).
\end{equation}
Next we analyze the inseparability of $\sigma_{b}$ for all possible bipartite splits namely $(AB)-C$, $A-(BC)$, $B-(AC)$.
\subsection{Bipartite split $(AB)-C$}
For the bipartite split $(AB)-C$, we have
\begin{equation}\label{}
    \sigma_{b}^{T_{C}}=\frac{1}{7b+1}
    \left(
       \begin{array}{cccccccc}
        b & 0 & 0 & 0 & 0 & 0 & 0 & b \\
    0 & b & 0 & 0 & b & 0 & 0 & 0 \\
    0 & 0 & b & 0 & 0 & 0 & 0 & 0 \\
    0 & 0 & 0 & b & 0 & 0 & b & 0 \\
    0 & b & 0 & 0 & \frac{1+b}{2} & \frac{\sqrt{1-b^{2}}}{2} & 0 & 0 \\
    0 & 0 & 0 & 0 & \frac{\sqrt{1-b^{2}}}{2} & \frac{1+b}{2} & 0 & 0 \\
    0 & 0 & 0 & b & 0 & 0 & b & 0 \\
    b & 0 & 0 & 0 & 0 & 0 & 0 & b \\
       \end{array}
     \right).
\end{equation}
It is easy to see that the state $\sigma_{b}^{T_{C}}$ is positive as
\begin{equation}\label{}
    \sigma_{b}^{T_{C}}=I\otimes I\otimes U\sigma_{b}I\otimes I \otimes U^{\dag},
\end{equation}
where
\begin{equation}\label{}
    U=\left(
        \begin{array}{cc}
          0 & 1 \\
          1 & 0 \\
        \end{array}
      \right).
\end{equation}
We now prove that $\sigma_{b}$ is an entangled state with respect to bipartite split $(AB)-C$ by using the range criterion.
Assume that $b\neq0,1$, then any vector belonging to the range of $\sigma_{b}$ can be presented as
\begin{equation}\label{}
    u=(A_{1},A_{2},A_{3},A_{4},xA_{5},A_{1}+A_{5},A_{2},A_{3}), \ \ A_{i}\in\mathcal{C}, i=1,\cdots,5,
\end{equation}
where $x=\sqrt{\frac{1+b}{1-b}}$.
On the one hand, for $x\neq0,1$, if $u$ is positive it must be of the form
\begin{equation}\label{}
    u_{prod}=(r,s,t,q)\otimes(\widetilde{A}_{1},\widetilde{A}_{2})
    =(r\widetilde{A}_{1},r\widetilde{A}_{2},s\widetilde{A}_{1},s\widetilde{A}_{2},
 t\widetilde{A}_{1},t\widetilde{A}_{2},q\widetilde{A}_{1},q\widetilde{A}_{2}),
\end{equation}
where $r,s,t,q,\widetilde{A}_{1},\widetilde{A}_{2}\in\mathcal{C}$.\\
Comparing the two forms of vector $u$, we consider the following cases.

(i) If $rs\neq0$, we can put $r=1,s=1$, then $A_{1}=A_{3}=\widetilde{A}_{1}$, $A_{2}=A_{4}=\widetilde{A}_{2}$,
$q\widetilde{A}_{1}=A_{2}$, $q\widetilde{A}_{2}=A_{3}$, and $(q^{2}-1)\widetilde{A}_{2}=0$. If $q^{2}\neq1$, then
$\widetilde{A}_{1}=\widetilde{A}_{2}=0$, $u=0$. If $q^{2}=1$, we put $q=1$, then $\widetilde{A}_{1}=\widetilde{A}_{2}$,
$xA_{5}=t\widetilde{A}_{1}$, $A_{1}+A_{5}=t\widetilde{A}_{1}$, and $t=\frac{x}{x-1}$.
We have
\begin{equation}\label{}
    u_{1}=A_{1}(1,1,\frac{x}{x-1},1)\otimes (1,1), \ \ A_{1}\in\mathcal{C}.
\end{equation}

(ii) If $r\neq0$, $s=0$, we put $r=1$, then $A_{1}=\widetilde{A}_{1}$, $A_{2}=\widetilde{A}_{2}$, $A_{3}=A_{4}=0$,
$q\widetilde{A}_{1}=A_{2}$, $q\widetilde{A}_{2}=0$. For the case $q\neq0$, we have $\widetilde{A}_{1}=\widetilde{A}_{2}$,
then $u=0$. For $q=0$, we get $A_{2}=\widetilde{A}_{2}=0$, $A_{5}=-A_{1}$, we get
\begin{equation}\label{}
    u_{2}=A_{1}(1,0,-x,0)\otimes(1,0), \ \ A_{1}\in\mathcal{C}.
\end{equation}

(iii) If $r=0$, $s\neq0$, we put $s=1$, then $A_{1}=A_{2}=q\widetilde{A}_{1}=0$,
$q\widetilde{A}_{2}=A_{3}=\widetilde{A}_{1}$. For $q\neq0$, we have $\widetilde{A}_{1}=\widetilde{A}_{2}=0$, then $u=0$.
For $q=0$, we get $\widetilde{A}_{1}=0$, $\widetilde{A}_{2}=A_{4}$, $A_{5}=t\widetilde{A}_{2}=0$, if $t\neq0$, then $\widetilde{A}_{2}=0$, $u=0$. Then we have
\begin{equation}\label{}
    u_{3}=A_{4}(0,1,0,0)\otimes(0,1), \ \ A_{4}\in\mathcal{C}.
\end{equation}

(iiii) If $r=0, s=0$, then $q\widetilde{A}_{1}=A_{2}=0$, $q\widetilde{A}_{2}=A_{3}=0$, $t\widetilde{A}_{1}=xA_{5}$,
$t\widetilde{A}_{2}=A_{5}$. For $q\neq0$, one has $\widetilde{A}_{1}=\widetilde{A}_{2}=0$, $u=0$. For $q=0$,
\begin{equation}\label{}
    u_{prod}=(0,0,0,0,t\widetilde{A}_{1},t\widetilde{A}_{2},0,0),
\end{equation}
if $t=0$, then $u=0$, we put $t=1$, then
\begin{equation}\label{}
    u_{4}=A_{5}(0,0,1,0)\otimes(x,1), \ \ A_{5}\in\mathcal{C}.
\end{equation}

All partial complex conjugations of vectors $u_{1}, u_{2}, u_{3}, u_{4}$ are
\begin{eqnarray}
 \nonumber u^{\star2}_{1} &=& A_{1}(1,1,\frac{x}{x-1},1)\otimes (1,1), \\
 \nonumber u^{\star2}_{2} &=& A_{1}(1,0,-x,0)\otimes(1,0), \\
 \nonumber u^{\star2}_{3} &=& A_{4}(0,1,0,0)\otimes(0,1), \\
  u^{\star2}_{4} &=& A_{5}(0,0,1,0)\otimes(x,1).
\end{eqnarray}

On the other hand, any vector belongs to the range of $\sigma_{b}^{T_{C}}$
can be written as
\begin{equation}\label{}
    u'=(A'_{1},A_{2}',A_{3}',A_{4}',A_{2}'+A_{5}',xA_{5}',A_{4}',A_{1}'), \ \ A'_{i}\in\mathcal{C}, i=1,\cdots,5.
\end{equation}

Let us check whether the vectors $u^{\star2}_{1}, u^{\star2}_{2}, u^{\star2}_{3}, u^{\star2}_{4}$
can be written in the above form.
For $u^{\star2}_{1}$, we obtain that $u^{\star2}_{1}$ belongs to the rang of $(\sigma_{b}^{T_{C}})$.
For $u^{\star2}_{2}$, assuming that it is of the form $u'$, we get $A_{1}=A'_{1}=0$, then $u^{\star2}_{2}$ is the trivial zero vector.
For $u^{\star2}_{3}$, we have $A_{4}=A'_{4}=0$, then $u^{\star2}_{3}$ is the trivial zero vector.
For $u^{\star2}_{4}$, considering $A'_{2}=0, A_{2}'+A_{5}'=xA_{5}$, $xA_{5}'=A_{5}$, we obtain $x^2=1$.
This contradicts the fact that $x=\sqrt{\frac{1+b}{1-b}}\neq0,1$.

In summery, for any $b\neq0,1$, the state $\sigma_{b}$ is a bound entangled state with respect to bipartite split $(AB)-C$.

\subsection{Bipartite split $A-(BC)$}
For the bipartite split $A-(BC)$, we have
\begin{equation}\label{}
   \sigma_{b}^{T_{BC}}=\frac{1}{7b+1}
   \left(
     \begin{array}{cccccccc}
    b & 0 & 0 & 0 & 0 & 0 & 0 & 0 \\
    0 & b & 0 & 0 & b & 0 & 0 & 0 \\
    0 & 0 & b & 0 & 0 & b & 0 & 0 \\
    0 & 0 & 0 & b & 0 & 0 & b & 0 \\
    0 & b & 0 & 0 & \frac{1+b}{2} & \frac{\sqrt{1-b^{2}}}{2} & 0 & 0 \\
    0 & 0 & b & 0 & \frac{\sqrt{1-b^{2}}}{2} & \frac{1+b}{2} & 0 & 0 \\
    0 & 0 & 0 & b & 0 & 0 & b & 0 \\
    0 & 0 & 0 & 0 & 0 & 0 & 0 & b \\
     \end{array}
   \right)
\end{equation}

For any nonzero real vector $X =(x_{1}, x_{2}, \cdots, x_{8})^{T} $, we have
\begin{eqnarray}
X^{T}\sigma_{b}^{T_{BC}}X&=&f(x_1, x_2, \ldots, x_{8})= bx_{1}^{2}+bx_{8}^{2}-bx_{5}^{2}+b(x_{2}+x_{5})^{2} \nonumber\\
&&+b(x_{3}+x_{6})^{2}+b(x_{4}+x_{7})^{2}+(\sqrt{1+b \over 2}x_{5}+ \sqrt{1-b \over 2}x_{6})^{2}.
\end{eqnarray}

Obviously, the positive index of inertia is $6$, and the rank of $\sigma_{b}^{T_{BC}}$ is $7$. Therefore $\sigma_{b}$
is an NPT state with respect to bipartite split $A-(BC)$.

Next we will show that the state $\sigma_{b}$ is 1-copy undistillable with respect to bipartite split $A-(BC)$. We begin with the following

\textbf{Theorem 1.}
A bipartite state $\rho$ acting on a Hilbert space $\mathcal{H}_{A}\otimes\mathcal{H}_{B}$ is distillable if and only if there exist a positive integer $N\in\mathbb{N}$ and a Schmidt rank-$2$ state vector $|\psi^{[N]}_{2}\rangle$
 in $\mathcal{H}^{\otimes N}_{A}\otimes\mathcal{H}^{\otimes N}_{B}$ such that \cite{x2}
\begin{equation}\label{}
    \langle\psi^{[N]}_{2}|(\rho^{\otimes N})^{T_{B}}|\psi^{[N]}_{2}\rangle=\langle\psi^{[N]}_{2}|(\rho^{T_{B}})^{\otimes N}|\psi^{[N]}_{2}\rangle<0.
\end{equation}

For $N=1$, the Schmidt rank-$2$ state is of the form
\begin{equation}\label{}
    |\psi^{[1]}_{2}\rangle=\sum^{2}_{k,i=1}\sum^{4}_{j=1}
    c_{k}u_{i}^{(k)}v_{j}^{(k)}|i\rangle_{A}\otimes|j\rangle_{BC},
\end{equation}
where $\sum^{2}_{k=1}c^2_{k}=1$, $\sum^{2}_{i=1}u_{i}^{(k_{1})*}u_{i}^{(k_{2})}=\delta_{k_{1}k_{2}}$,
$\sum^{4}_{j=1}v_{j}^{(k_{1})*}v_{j}^{(k_{2})}=\delta_{k_{1}k_{2}}$.
So we have
\begin{equation}\label{nj}
    \langle\psi^{[1]}_{2}|\sigma_{b}^{T_{BC}}|\psi^{[1]}_{2}\rangle
    =\sum_{k_{1},k_{2},i=1}^2\sum^{4}_{j=1}\frac{1}{7b+1}
    c^{*}_{k_{1}}c_{k_{2}}u_{i}^{(k_{1})*}(M_{(k_{1},k_{2})})_{i,j}u_{i}^{(k_{2})}
    =\frac{1}{7b+1}Y_{1}^{\dag}M_{1}Y_{1}
\end{equation}
with $Y_{1}=(c_{1}u^{1}_{1},c_{1}u^{1}_{2},c_{2}u^{2}_{1},c_{2}u^{2}_{2})^{T}$. We get the
matrix $M_{1}$ is positive. According to the Theorem $1$, the state $\sigma_{b}$ is 1-copy undistillable with respect to bipartite split $A-(BC)$.

\subsection{Bipartite split $B-(AC)$}
For the bipartite split $B-(AC)$, we can use the same method as above, for any nonzero real vector $X =(x_{1}, x_{2}, \cdots, x_{8})^{T} $, we have
\begin{eqnarray}
  X^{T}\sigma_{b}^{T_{AC}}X&=& f(x_1, x_2, \ldots, x_{8})= bx_{2}^{2}+bx_{7}^{2}-bx_{5}^{2}+b(x_{1}+x_{6})^{2} \nonumber\\
&&+b(x_{3}+x_{8})^{2}+b(x_{4}+x_{5})^{2}+(\sqrt{1+b \over 2}x_{5}+ \sqrt{1-b \over 2}x_{6})^{2}.
\end{eqnarray}

The positive index of inertia is $6$, and the rank of $\sigma_{b}^{T_{AC}}$ is $7$, then $\sigma_{b}$
is also a NPT state with respect to bipartite split $B-(AC)$.

In the similar way, by direct calculation we have $\langle\psi^{[1]}_{2}|(\sigma_{b}^{T_{AC}}|\psi^{[1]}_{2}\rangle\geq 0$ for all the
Schmidt rank-$2$ states $|\psi^{[1]}_{2}\rangle$ in $\mathcal{H}^{\otimes 1}_{B}\otimes\mathcal{H}^{\otimes 1}_{AC}$. Therefore, $\sigma_b$ is 1-copy undistillable with respect to bipartite split $B-(AC)$.

\section{Entanglement of $2n\otimes2n\otimes2n$ quantum systems}
Consider the Hilbert space $\mathcal{H}=\mathcal{H}_{A}\otimes\mathcal{H}_{B}\otimes\mathcal{H}_{C}$, $dim\mathcal{H}_{A}=dim\mathcal{H}_{B}=dim\mathcal{H}_{C}=2n$. Let $P_{\phi}=|\phi\rangle\langle\phi|$, ${e_{i}}$ stand for orthonormal basis of $\mathcal{C}^{2n}$, $i=1,2,\cdots,2n$. We define the vectors
\begin{eqnarray}
  \Psi_{ijk} &=& \frac{1}{\sqrt{2}}(e_{i}\otimes e_{j}\otimes e_{k}+e_{n+i}\otimes e_{j}\otimes e_{k+1}), \nonumber \\
  \Psi_{ik} &=& \frac{1}{\sqrt{2}}(e_{i}\otimes e_{k}\otimes e_{2n}+e_{n+i}\otimes e_{k+1}\otimes e_{1}), \nonumber \\
  \Phi_{a}&=&e_{n+1}\otimes e_{1}\otimes(\sqrt{\frac{1+a}{2}}e_{1}+\sqrt{\frac{1-a}{2}}e_{2n}), \ \ a\in[0,1].
\end{eqnarray}
where $i=1,\cdots,n$, $j=1,\cdots,2n$, $k=1,\cdots,2n-1$.
Now we define the following state
\begin{equation}\label{}
    \rho_{insep}=\frac{2}{8n^3-1}\sum_{i=1}^n\sum_{j=1}^{2n}\sum_{k=1}^{2n-1}(P_{\Psi_{ijk}}+P_{\Psi_{ik}})+
    \frac{1}{8n^3-1}P_{e_{n}\otimes e_{2n}\otimes e_{n}}.
\end{equation}

This state is inseparable with respect to all bipartite splits as there always exist a minor matrix of order $2$ of its partial transposition is negative.
Mixing the states $\rho_{insep}$ and $P_{\Phi_{a}}$, we have
\begin{equation}\label{}
    \rho_{a}=\frac{(8n^3-1)a}{(8n^3-1)a+1}\rho_{insep}+\frac{1}{(8n^3-1)a+1}P_{\Phi_{a}}.
\end{equation}

Next we analyze the different types of entanglement of $\rho_{a}$ for all possible bipartite splits.

\subsection{Bipartite split $(AB)-C$}
For the bipartite split $(AB)-C$, $\rho_{a}^{T_{C}}$ is a $4n^2\times 4n^2$ matrix
\begin{equation}\label{}
    \rho_{a}^{T_{C}}=\frac{1}{(8n^3-1)a+1}\left(
                                    \begin{array}{cccccccc}
                                      F_{1} & 0 & \cdots & 0 & G^{t}_{1} & H^{t}_{1} & \cdots & 0 \\
                                      0 & F_{1} & \cdots & 0 & 0 & G^{t}_{1} & \cdots & 0 \\
                                       \vdots & \vdots & \cdots & \vdots & \vdots & \vdots & \cdots & \vdots  \\
                                      0 & 0 & \cdots & F_{1} & 0 & 0 & \cdots & G^{t}_{1} \\
                                      G_{1} & 0 & \cdots & 0 & K_{1} & 0 & \cdots & 0 \\
                                      H_{1} & G_{1} & \cdots & 0 & 0 & F_{1} & \cdots & 0 \\
                                      \vdots & \vdots & \cdots & \vdots & \vdots & \vdots & \cdots & \vdots \\
                                      0 & 0 & \cdots & G_{1} & 0 & 0 & \cdots & F_{1} \\
                                    \end{array}
                                  \right),
\end{equation}
with
$$
F_{1}=\left(
        \begin{array}{cccc}
          a & 0 & \cdots & 0 \\
          0 & a & \cdots & 0 \\
          \vdots & \vdots & \cdots & \vdots \\
          0 & 0 & \cdots & a \\
        \end{array}
      \right), \ \
 G_{1}=\left(
         \begin{array}{ccccc}
           0 & a & 0 & \cdots & 0 \\
           0 & 0 & a & \cdots & 0 \\
           \vdots & \vdots & \vdots & \cdots & \vdots \\
           0 & 0 & 0 & \cdots & a \\
           0 & 0 & 0 & \cdots & 0 \\
         \end{array}
       \right), $$

$$
H_{1}=\left(
        \begin{array}{cccc}
          0 & 0 & 0 & 0 \\
          \vdots & \vdots & \cdots & \vdots \\
          0 & 0 & \cdots & 0 \\
          a & 0 & \cdots & 0 \\
        \end{array}
      \right), \ \
 K_{1}=\left(
         \begin{array}{ccccc}
           \frac{1+a}{2} & 0 & \cdots & 0 & \frac{\sqrt{1-a^2}}{2} \\
           0 & a & \cdots & 0 & 0 \\
           \vdots & \vdots & \cdots & \vdots & \vdots \\
           0 & 0 & \cdots & a & 0 \\
           \frac{\sqrt{1-a^2}}{2} & 0 & \cdots & 0 & \frac{1+a}{2} \\
         \end{array}
       \right),
      $$
where $F_{1},G_{1},H_{1},K_{1}$ are all
$2n\times 2n$ matrices and $G^{t}$ stand for transposition of $G$.\\
For any nonzero real vector $X=(x_1,x_2,\cdots,x_{8n^3})^T$, we get
\begin{eqnarray}
  X^{T}\rho_{a}^{T_{C}}X &=& \sum^{2n^2-1}_{k=0}\sum^{2n}_{i=2} a(x_{i+2nk}+x_{4n^3+i+2nk-1})^{2}+ \sum^{2n^2-2}_{k=0}a(x_{1+2nk}+x_{4n^3+4n+2nk})^{2} \nonumber \\
  && +ax_{4n^3-2n+1}^{2}+(\sqrt{1-a \over 2}x_{4n^3+1}+ \sqrt{1+a \over 2}x_{4n^3+2n})^{2}.
\end{eqnarray}
Obviously, the positive index of inertia is $4n^3+1$, and the rank of
$\rho_a^{T_C}$ is $4n^3+1$. We drive that the state $\rho_a^{T_C}$ is a PPT state.

Next, we will show that the state $\rho_a$ is entangled with respect to bipartite split $(AB)-C$.
For any vector belongs to the range of $\rho_a^{T_C}$ can be presented as
\begin{eqnarray}
 && v = (A_{1},A_{2},\cdots,A_{2n-1},A_{2n},A_{2n+1},\cdots,A_{4n-1},A_{4n},\cdots,A_{4n^3-2n+1},\cdots,A_{4n^3-1},A_{4n^3}, \nonumber \\
   && A_{2}+B,A_{3},\cdots,A_{2n},yB,A_{2n+2},\cdots,A_{4n},A_{1},\cdots,A_{4n^3-2n+2},\cdots,A_{4n^3},A_{4n^3-4n+1}),
\end{eqnarray}
where $y=\sqrt{\frac{1+a}{1-a}}$, $A_{i},B\in\mathcal{C}, i=1,2,\cdots,4n^3$.

For $y\neq0,1$, if $v$ is positive, it must be of the form
\begin{equation}\label{}
    v_{prod}=(s_{1},s_{2},\cdots,s_{4n^2})\otimes (\widetilde{A}_{1},\widetilde{A}_{2},\cdots,\widetilde{A}_{2n}),\ \ \
    s_{i},\widetilde{A}_{j}\in\mathcal{C}, i=1,\cdots,4n^2, j=1,\cdots,2n.
\end{equation}
Let us now consider the following cases, comparing the two forms of vector $v$.\\
(i) While $s_{1}=0$, we have $s_{m}=s_{m+2n^2}=0$ and $s_{4n^2}=0$, $m=2,3,\cdots,2n^2-1$. The proof is in Appendix A.
Hence if $s_{2n^2}=0$, then $s_{2n^2+1}\neq0$, otherwise $v=0$, we can put $s_{2n^2+1}=1$, then we get
\begin{equation}\label{}
    v_{1}=B(0,0,\cdots,0,1,0,\cdots,0)\otimes (1,0,\cdots,0,y).
\end{equation}
If $s_{2n^2}\neq0$, combine with $s_{4n^2}(\widetilde{A}_{1},\cdots,\widetilde{A}_{2n-1})=s_{2n^2}(\widetilde{A}_{2},\cdots,\widetilde{A}_{2n})$
and $\frac{s_{2n^2+1}}{x}\widetilde{A}_{2n}=s_{2n^2+1}\widetilde{A}_{1}$
one has $s_{2n^2+1}=0$, we put $s_{2n^2}=1$, so we get
\begin{equation}\label{}
    v_{2}=A_{4n^3-2n+1}(0,0,\cdots,1,0,0,\cdots,0)\otimes (1,0,\cdots,0,0).
\end{equation}
(ii) While $s_{1}\neq0$, we put $s_{1}=1$, then $\widetilde{A}_{i}=A_{i}$, $i=1,2,\cdots,2n$. According to the relation
$A_{k}=s_{2n^2+1}A_{k-1}$, $3\leq k\leq2n$, we have that if for some $k$, $A_k\neq0$, $2\leq k\leq2n$, then $A_{2},\cdots,A_{2n}$ are not zero and $s_{2n^2+1}\neq0$, if for some $k$, $A_{k}=0$, $2\leq k\leq2n$, then
$A_{2}=\cdots=A_{2n}=0$, $s_{2n^2+1}\neq0$.\\
If $A_{1}=0$, from $A_{1}=s_{2n^2+2}A_{2n}$, then $s_{2n^2+2}=0$, $A_{2n}\neq0$, otherwise $v=0$, according to the conclusion of Appendix A and $s_{2n^2}A_{2n}=s_{4n^2}A_{2n-1}$, one has $s_{2n^2}=0$, therefore
\begin{equation}\label{}
    v_{3}=A_{2}(1,0,\cdots,0,s_{2n^2+1},0,\cdots,0)\otimes(0,1,s_{2n^2+1},s^2_{2n^2+1},\cdots,s^{2n-2}_{2n^2+1}).
\end{equation}
If $A_{1}\neq0$, we put $s_{2n^2+1}=1$, then $A_{2}+B=A_{1}$, $yB=A_{2n}$ and $A_{2}=,\cdots,=A_{2n}$. From $A_{1}=s_{2n^2+2}A_{2n}$, we obtain $s_{2n^2+2}=\frac{y+1}{y}$. Since $s_{m}A_{2}=s_{2n^2+m}A_{1}$, $2\leq m\leq2n^2$ and $s_{m}A_{1}=s_{2n^2+m+1}A_{2n}$, $2\leq m\leq2n^2-1$, then $s_{m}=(\frac{y+1}{y})^{2m-2}$, $s_{2n^2+m}=(\frac{y+1}{y})^{2m-3}$, we have
\begin{eqnarray}
  v_{4} &=& A_{2}(1,(\frac{y+1}{y})^{2},(\frac{y+1}{y})^{4},\cdots,(\frac{y+1}{y})^{4n^2-2},1,\frac{y+1}{y}
    ,\cdots,(\frac{y+1}{y})^{4n^2-3}) \nonumber \\
  && \otimes (\frac{y+1}{y},1,1,\cdots,1).
\end{eqnarray}

All partial complex conjugations of vectors $v_{1}, v_{2}, v_{3}, v_{4}$ are
\begin{eqnarray}
  v^{\star2}_{1}&=&B(0,0,\cdots,0,1,0,\cdots,0)\otimes (1,0,\cdots,0,y), \nonumber \\
  v^{\star2}_{2}&=&A_{4n^3-2n+1}(0,0,\cdots,1,0,0,\cdots,0)\otimes (1,0,\cdots,0,0), \nonumber \\
  v^{\star2}_{3}&=&A_{2}(1,0,\cdots,0,s_{2n^2+1},0,\cdots,0)\otimes(0,1,{s_{2n^2+1}}^*,{s^2_{2n^2+1}}^*,
  \cdots,{s^{2n-2}_{2n^2+1}}^*), \ \ \ s_{2n^2+1}\neq0,\nonumber \\
  v^{\star2}_{4} &=& A_{2}(1,(\frac{y+1}{y})^{2},(\frac{y+1}{y})^{4},\cdots,(\frac{y+1}{y})^{4n^2-2},1,\frac{y+1}{y}
    ,\cdots,(\frac{y+1}{y})^{4n^2-3}) \nonumber \\
  && \otimes (\frac{y+1}{y},1,1,\cdots,1).
\end{eqnarray}

On the other hand, any vector belongs to the range of $\rho_{a}$
can be written as
\begin{eqnarray}
 && v' = (A'_{1},A'_{2},\cdots,A'_{2n},\cdots,A'_{4n^3-2n+1},A'_{4n^3-2n+2},\cdots,A'_{4n^3},yB',A'_{1},\cdots, A'_{2n-2},\nonumber \\
   &&B'+A'_{2n-1},A'_{2n},A'_{2n+1},\cdots,A'_{4n-1},\cdots,A'_{4n^3-2n},A'_{4n^3-2n+1},\cdots,A'_{4n^3-1}),
\end{eqnarray}

Now we check whether vectors $v^{\star2}_{1}, v^{\star2}_{2}, v^{\star2}_{3}, v^{\star2}_{4}$
can be written in the above form.

For $v^{\star2}_{1}$, assume it can be written as the form of $v'$, we get $B=yB'$, $yB=B'$, then $y^2=1$, which contradicts the fact that $y\neq0,1$. For $v^{\star2}_{2}$, we certainly have $A_{4n^3-2n+1}=A'_{4n^3-2n+1}=0$, then $v^{\star2}_{2}$ is the zero vector.For $v^{\star2}_{3}$, it must be hold $s_{2n^2+1}A_{2}=A'_{1}=0$, then $s_{2n^2+1}=0$. This contradicts the fact that
$s_{2n^2+1}\neq0$. For $v^{\star2}_{4}$, considering $A'_{1}=\frac{y+1}{y}A_{2}$ and $A'_{1}=A_{2}$, we get $A_{2}=0$, then $v^{\star2}_{4}$ is also the zero vector.

Therefore, it leads to the conclusion that none of vectors $v^{\star2}_{1}, v^{\star2}_{2}, v^{\star2}_{3}, v^{\star2}_{4}$ belongs to the range of $\rho_{a}$. For any $a\neq0,1$, the state $\rho_a$ is a bound entangled state with respect to bipartite split $(AB)-C$.

\subsection{Bipartite split $A-(BC)$}
For the bipartite split $A-(BC)$, we have $\rho_{a}^{T_{BC}}$ is a $2n\times2n$ matrix
\begin{equation}\label{}
   \rho_{a}^{T_{BC}}=\frac{1}{(8n^3-1)a+1}\left(
     \begin{array}{cccccccccc}
       F_2 & 0 & 0 & \cdots & 0 & G'_2 & H'_2 & 0 & \cdots & 0 \\
       0 & F_2 & 0 & \cdots & 0 & 0 & G'_2 & H'_2 & \cdots & 0 \\
       \vdots & \vdots & \vdots & \cdots & \vdots & \vdots & \vdots & \vdots & \cdots & \vdots \\
       0 & 0 & 0 & \cdots & F_2 & 0 & 0 & 0 & \cdots & G'_2 \\
       G_2 & 0 & 0 & \cdots & 0 & K_2 & 0 & 0 & \cdots & 0 \\
       H_2 & G_2 & 0 & \cdots & 0 & 0 & F_2 & 0 & \cdots & 0 \\
       \vdots & \vdots & \vdots & \cdots & \vdots & \vdots & \vdots & \vdots & \cdots & \vdots \\
       0 & 0 & 0 & \cdots & G_2 & 0 & 0 & 0 & \cdots & F_2 \\
     \end{array}
   \right)
\end{equation}
with
$$
F_{2}=\left(
        \begin{array}{cccc}
          a & 0 & \cdots & 0 \\
          0 & a & \cdots & 0 \\
          \vdots & \vdots & \cdots & \vdots \\
          0 & 0 & \cdots & a \\
        \end{array}
      \right), \ \
 G_{2}=\left(
         \begin{array}{ccccc}
           0 & a & 0 & \cdots & 0 \\
           0 & 0 & a & \cdots & 0 \\
           \vdots & \vdots & \vdots & \cdots & \vdots \\
           0 & 0 & 0 & \cdots & a \\
           0 & 0 & 0 & \cdots & 0 \\
         \end{array}
       \right), \ \
H_{2}=\left(
        \begin{array}{cccc}
          0 & 0 & 0 & 0 \\
          \vdots & \vdots & \cdots & \vdots \\
          0 & 0 & \cdots & 0 \\
          a & 0 & \cdots & 0 \\
        \end{array}
      \right),
$$
and
$$
K_2=\left(
  \begin{array}{cccccccc}
    \frac{1+a}{2} & 0 & \cdots & 0 & \frac{\sqrt{1-a^2}}{2} & 0 & \cdots & 0 \\
    0 & a & \cdots & 0 & 0  & 0 & \cdots & 0 \\
    \vdots & \vdots & \cdots & \vdots & \vdots & \vdots & \cdots & \vdots \\
    0 & 0 & \cdots & a & 0 &  0 & \cdots & 0 \\
    \frac{\sqrt{1-a^2}}{2} & 0 & \cdots & 0 & \frac{1+a}{2} &  0 & \cdots & 0 \\
    0 & 0 & \cdots & 0 & 0 &  a & \cdots & 0 \\
    \vdots & \vdots & \cdots & \vdots & \vdots & \vdots & \cdots & \vdots \\
    0 & 0 & \cdots & 0 & 0 & 0 & \cdots & a \\
  \end{array}
\right),
$$
which are $4n^2\times 4n^2$ matrices.

For any nonzero real vector $X=(x_1,x_2,\cdots,x_{8n^3})^T$, we get
\begin{eqnarray}
  &&X^{T}\rho_{a}^{T_{BC}}X = \sum^{n-1}_{k=0}\sum^{4n^2}_{i=2} a(x_{i+4kn^2}+x_{4n^3+4kn^2+i-1})^{2}+ \sum^{n-2}_{k=0}a(x_{1+4kn^2}+x_{4n^3+(8+4k)n^2})^{2} \nonumber \\
  && +(\sqrt{1-a \over 2}x_{4n^3+2n}+ \sqrt{1+a \over 2}x_{4n^3+1})^{2}+ax_{4n^3-4n^2+1}^{2}+ax_{4n^3+4n^2}^{2}-ax_{4n^3+1}^{2}.
\end{eqnarray}
Obviously, the state $\rho_{a}^{T_{BC}}$ is not positive, so $\rho_{a}$ is a NPT state with respect to the bipartite split $A-(BC)$.

Now we prove $\rho_{a}$ is 1-copy undistillable with respect to the bipartite split $A-(BC)$ by using Theorem $1$.

For $N=1$, the Schmidt rank-$2$ state is of the form
\begin{equation}\label{}
    |\varphi^{[1]}_{2}\rangle=\sum^{2}_{k=1}\sum^{2n}_{i=1}\sum^{4n^2}_{j=1}
    c_{k}u_{i}^{(k)}v_{j}^{(k)}|i\rangle_{A}\otimes|j\rangle_{BC},
\end{equation}
where $\sum^{2}_{k=1}c^2_{k}=1$, $\sum^{2n}_{i=1}u_{i}^{(k_{1})*}u_{i}^{(k_{2})}=\delta_{k_{1}k_{2}}$,
$\sum^{4n^2}_{j=1}v_{j}^{(k_{1})*}v_{j}^{(k_{2})}=\delta_{k_{1}k_{2}}$.
Then we have
\begin{eqnarray}
  \langle\varphi^{[1]}_{2}|\rho_{a}^{T_{BC}}|\varphi^{[1]}_{2}\rangle
  &=&\sum_{k_{1},k_{2}=1}^2\sum^{2n}_{i=1}\sum^{4n^2}_{j=1}\frac{1}{(8n^3-1)a+1}
    c^{*}_{k_{1}}c_{k_{2}}u_{i}^{(k_{1})*}(M_{(k_{1},k_{2})})_{i,j}u_{i}^{(k_{2})}\nonumber \\
&&= \frac{1}{(8n^3-1)a+1}Y_{2}^{\dag}M_{2}Y_{2}
\end{eqnarray}
with $Y_{2}=(c_{1}u^{1}_{1},c_{1}u^{1}_{2},\cdots,c_{1}u^{1}_{2n},c_{2}u^{2}_{1},c_{2}u^{2}_{2},\cdots,c_{2}u^{2}_{2n})^{T}$, and the matrix $M_{2}$ are positive, that is $\langle\varphi^{[1]}_{2}|\rho_{a}^{T_{BC}}|\varphi^{[1]}_{2}\rangle\geq0$ for any Schmidt rank-$2$ state vector $|\varphi^{[1]}_{2}\rangle$ in $\mathcal{H}^{\otimes 1}_{A}\otimes\mathcal{H}^{\otimes 1}_{BC}$. Therefore $\rho_{a}$ is 1-copy undistillable with respect to the bipartite split $A-(BC)$.

\subsection{Bipartite split $B-(AC)$}
We can use the same method to analyze the case of bipartite split $B-(AC)$. For any nonzero real vector $X=(x_1,x_2,\cdots,x_{8n^3})^T$, we get
\begin{eqnarray}
  &&X^{T}\rho_{a}^{T_{AC}}X = \sum^{2n^2-1}_{k=0}\sum^{2n-1}_{i=1} a(x_{i+2kn}+x_{4n^3+2kn+i-1})^{2}+ \sum^{2n^2-2}_{k=0}a(x_{2n(2+k)}+x_{4n^3+2kn+1})^{2} \nonumber \\
  && +(\sqrt{1-a \over 2}x_{4n^3+2n}+ \sqrt{1+a \over 2}x_{4n^3+1})^{2}+ax_{4n^3+2n+1}^{2}+ax_{2n}^{2}-ax_{4n^3+1}^{2},
\end{eqnarray}
then $\rho_{a}^{T_{AC}}$ is not positive, $\rho_{a}$ is a NPT state with respect to the bipartite split $B-(AC)$.
By direct calculation $\langle\varphi^{[1]}_{2}|\rho_{a}^{T_{AC}}|\varphi^{[1]}_{2}\rangle$ is positive, where $\varphi^{[1]}_{2}\in\mathcal{H}^{\otimes 1}_{B}\otimes\mathcal{H}^{\otimes 1}_{AC}$.

Therefore $\rho_{a}$ is 1-copy undistillable with respect to bipartite split $B-(AC)$.

\section{Conclusion and Discussion}

In summary, we have constructed a class of tripartite entangled states, then presented a detailed description about the entanglement with respect to all possible bipartite splits in $2\otimes2\otimes2$ systems. For the bipartite split$(AB)-C$, the state is bound entanglement, for another two bipartite splits, it is a NPT state and 1-copy undistillable. Finally, we have generalized the results to the case of $2n\otimes2n\otimes2n$ systems.

In order to avoid complicated calculations, we can also use the following method to prove 1-copy undistillation. According to the Ref. \cite{sss}, a bipartite state $\rho$ acting on a Hilbert space $\mathcal{H}_{A}\otimes\mathcal{H}_{B}$ is distillable if and only if there exist a positive integer $K$ and two $2$-dimensional projectors $P$ : $(\mathcal{H}_{A})^{\otimes K}\longrightarrow \mathbb{C}^2$ and $Q$ : $(\mathcal{H}_{B})^{\otimes K}\longrightarrow \mathbb{C}^2$ such that $((P\otimes Q)\rho^{\otimes K}(P\otimes Q))^{T_{B}}$ is not positive.
For example, in $2\otimes2\otimes2$ systems, let $\{|1\rangle, |2\rangle\}$ and $\{|1\rangle, |2\rangle, |3\rangle, |4\rangle\}$ be orthonormal bases of $\mathcal{H}_{A}$ and $\mathcal{H}_{BC}$ respectively, we take $K=1$, considering the following two-dimensional projectors $P=|1\rangle\langle1|+|2\rangle\langle2|$ and $Q_{1}=|1\rangle\langle1|+|2\rangle\langle2|$. Then the nonzero eigenvalues of matrix $((P\otimes Q_{1})\sigma_{b}(P\otimes Q_{1}))^{T_{BC}}$ are $\frac{b}{7b+1}, \frac{1}{7b+1}(b\pm\frac{\sqrt{2b^2-2b+1}}{2}+\frac{1}{2})$, which are positive for $b\in(0,1)$. For another possible two-dimensional projectors $Q_{i}$ of $\mathcal{H}_{BC}$, $i=2,3,\cdots,6$, we also get the matrix $((P\otimes Q_{i})\sigma_{b}(P\otimes Q_{i}))^{T_{BC}}$ is positive by calculating the eigenvalues, then $\sigma_{b}$ is 1-copy undistillable with respect to the bipartite split $A-(BC)$. Using the same method, it is also easy to get $\sigma_{a}$ is 1-copy undistillable with respect to the bipartite splits $B-(AC)$.

In $2n\otimes2n\otimes2n$ systems, let $K=1$, according to the form of matrix $\rho_{a}^{T_{BC}}$, after taking every possible
two-dimensional projectors $P$ and $Q$ of $\mathcal{H}_{A}$ and $\mathcal{H}_{BC}$ respectively, the nonzero
rows and columns of matrix $(P\otimes Q)\rho_{a}^{T_{BC}}(P\otimes Q)$ constituting a new matrix $J$
only has five kinds of form as following
\begin{equation}\label{f}
    \left(
       \begin{array}{cccc}
         a & 0 & 0 & 0 \\
         0 & a & 0 & 0 \\
         0 & 0 & a & 0 \\
         0 & 0 & 0 & a \\
       \end{array}
     \right),
       \left(                                        \begin{array}{cccc}
                                                 \frac{1+a}{2} & 0 & 0 & 0 \\
                                                 0 & a & 0 & 0 \\
                                                 0 & 0 & a & 0 \\
                                                 0 & 0 & 0 & a \\
                                               \end{array}
                                             \right),
                                             \left(
                           \begin{array}{cccc}
                             a & 0 & 0 & a \\
                             0 & a & a & 0 \\
                             0 & a & a & 0 \\
                             0 & 0 & 0 & a \\
                           \end{array}
                         \right),
                                             \left(
                                                                   \begin{array}{cccc}
                                                                     a & 0 & 0 & 0 \\
                                                                     0 & a & a & 0 \\
                                                                     0 & a & \frac{1+a}{2} & 0 \\
                                                                     0 & 0 & 0 & a \\
                                                                   \end{array}
                                                                 \right),
                                                                 \left(
                                                                                    \begin{array}{cccc}
                                                                                      a & 0 & 0 & 0 \\
                                                                                      0 & a & 0 & 0 \\
                                                                                      0 & 0 & \frac{1+a}{2} & \frac{\sqrt{1-a^2}}{2} \\
                                                                                      0 & 0 & \frac{\sqrt{1-a^2}}{2} & \frac{1+a}{2} \\
                                                                                    \end{array}
                                                                                  \right).
\end{equation}
Obviously, the nonzero eigenvalues of matrix $(P\otimes Q)\rho_{a}^{T_{BC}}(P\otimes Q)$ are equal to the one of matrix $J$. It is easy to check that all eigenvalues of $J$ are positive for $a\in(0,1)$, then
$(P\otimes Q)\rho_{a}^{T_{BC}}(P\otimes Q)$ is positive for all two-dimensional projectors $P$ and $Q$. Therefore, $\rho_{a}$ is $1$-copy undistillable with respect to the bipartite split $A-(BC)$. Using the same method to analyze the case of bipartite split $B-(AC)$, we get $\rho_{a}$ is also $1$-copy undistillable.

We also hope that our results will help further investigations of multipartite quantum systems.

\appendix
\section{}
Comparing the two forms of $v$, we have
\begin{equation}\label{a1}
    s_m(\widetilde{A}_2,\widetilde{A}_3,\cdots,\widetilde{A}_{2n})
    =s_{2n^2+m}(\widetilde{A}_1,\widetilde{A}_2,\cdots,\widetilde{A}_{2n-1}),
\end{equation}
\begin{equation}\label{a2}
    s_{m-1}\widetilde{A}_1=s_{2n^2+m}\widetilde{A}_{2n},
\end{equation}
where $m=2,3,\cdots,2n^2$.
In fact, we can obtain the two relations from (\ref{a1}) and (\ref{a2}),\\
(i) if $s_{m-1}=0$, then $s_{2n^2+m}=0$, $m=2,3,\cdots,2n^2$,\\
(ii) if $s_{2n^2+m}=0$, then $s_{m}=0$, $m=2,3,\cdots,2n^2-1$.\\
Let us prove the first one. Assume that $s_{m-1}=0$ and $s_{2n^2+m}\neq0$, $m=2,3,\cdots,2n^2$. From (\ref{a2}), we have
$\widetilde{A}_{2n}=0$, then $s_{m}\widetilde{A}_1=s_{2n^2+m+1}\widetilde{A}_{2n}=0$. Here, if $s_{m}\neq0$, the $\widetilde{A}_1=0$, according to (\ref{a1}), we get $\widetilde{A}_2=0$, $\widetilde{A}_3=0$, $\cdots$, $\widetilde{A}_{2n}=0$, $v$ is a zero vector, so $s_{m}=0$. From
(\ref{a1}) and $s_{2n^2+m}\neq0$. $\widetilde{A}_1=0$, $\widetilde{A}_2=0$, $\cdots$, $\widetilde{A}_{2n-1}=0$ must hold , $v$ is also a zero vector, so $s_{2n^2+m}=0$.\\
For the second one, if $s_{2n^2+m}=0$, and $s_{m}\neq0$, $m=2,3,\cdots,2n^2-1$, then $\widetilde{A}_2=0$, $\widetilde{A}_3=0$, $\cdots$, $\widetilde{A}_{2n}=0$. Since $s_{m}\widetilde{A}_1=s_{2n^2+m+1}\widetilde{A}_{2n}$,
we get $\widetilde{A}_1=0$, then $v=0$. Therefore if $s_{2n^2+m}=0$, then $s_{m}=0$.

\begin{acknowledgements}
This work is supported by the China Scholarship Council, the National Natural Science Foundation of
China (11271138, and 11275131), Beijing Natural Science Foundation
Program and Scientific Research Key Program of Beijing Municipal
Commission of Education (KZ201210028032) and the Importation and Development of High-Caliber Talents Project of Beijing Municipal Institutions (CITTCD201404067).
\end{acknowledgements}


\begin{thebibliography}{24}
\bibitem{1}  C.H. Bennett and S.J. Wiesner, Phys. Rev. Lett. \textbf{69}, 2881 (1992).
\bibitem{2}  C.H. Bennett, G. Brassard, C. Cr¡äepeau, R. Jozsa, A. Peres and W.K. Wootters, Phys.
Rev. Lett. \textbf{70}, 1895 (1993).
\bibitem{3}  A. Ekert, Phys. Rev. Lett. \textbf{67}, 661 (1991).
\bibitem{4}  D. Deutsch, A. Ekert, R. Jozsa, C. Macchiavello, S. Popescu and A. Sanpera, Phys. Rev.
Lett. \textbf{77}, 2818 (1996).
\bibitem{5}  C.A. Fuchs, N. Gisin, R.B. Griffths, C-S. Niu and A. Peres, Phys. Rev. A \textbf{56}, 1163 (1997).
\bibitem{6}  R. F. Werner, Phys. Rev. A \textbf{40}, 4277 (1989).
\bibitem{t1} R. Horodecki, P. Horodecki, M. Horodecki and K. Horodecki, Rev. Mod. Phys. \textbf{81}, 865 (2009).
\bibitem{t2} H. Zhao, X.H. Zhang, S.M. Fei and Z.X. Wang, Chin. Sci. Bull. \textbf{58}, 2334 (2013).
\bibitem{t3} N. Brunner, J. Sharam and T. Vertesi, Phys. Rev. Lett. \textbf{108}, 110501 (2012).
\bibitem{t4} S.Q. Yan, Y. Guo and J.C. Hou, Chin. Sci. Bull. \textbf{59}, 279 (2014).
\bibitem{t5} W. Wen, Sci. China. Phys. Mech. Astron, \textbf{56}, 974 (2013).
\bibitem{t6} Y.Z. Wang, J.C. Hou and Y. Guo, Chin. Sci. Bull. \textbf{57}, 1643 (2012).
\bibitem{x1} A. Peres Phys. Rev. Lett. \textbf{77}, 1413 (1996).
\bibitem{j1}  P. Horodecki, Phys. Lett. A \textbf{232}, 333 (1997).
\bibitem{x2}  M. Horodecki, P. Horodecki and R. Horodecki, Phys.
Rev. Lett. \textbf{80}, 5239 (1998).

\bibitem{x3}  M. Horodecki, P. Horodecki and R. Horodecki, Phys. Rev. Lett. \textbf{78}, 574 (1997).
\bibitem{j2}  M. Horodecki and P. Horodecki, Phys. Rev. A \textbf{59}, 4206 (1999).
\bibitem{j3}  D. P. DiVincenzo, P. W. Shor, J. A. Smolin, B. M. Terhal and A. V.
Thapliyal, Phys. Rev. A \textbf{61}, 062312 (2000).
\bibitem{j4}  W. D¨¹r, J. I. Cirac, M. Lewenstein and D. Bru{\ss}, Phys. Rev. A \textbf{61}, 062313 (2000).
\bibitem{j5}  T. Eggeling, K. G. H. Vollbrecht, R. F. Werner and M. M. Wolf,
Phys. Rev. Lett. \textbf{87}, 257902 (2001).
\bibitem{j6}  B. Kraus, M. Lewenstein and J. I. Cirac, Phys. Rev. A, \textbf{65}, 042327 (2002).
\bibitem{j7}  R. O. Vianna and A. C. Doherty, Phys. Rev. A \textbf{74}, 052306 (2006).
\bibitem{j8}  P. Horodecki, M. Horodecki and R. Horodecki, Phys.
Rev. Lett. \textbf{82}, 1056 (1999).
\bibitem{j9}  P. W. Shor, J. A. Smolin and B. M. Terhal, Phys. Rev. Lett. \textbf{86}, 2681 (2001).
\bibitem{sss}S. L. Braunstein, S. Ghosh and S. Severini, Ann. Comb. \textbf{10}, 3 (2006).


\end{thebibliography}

\end{document}